# Ethical Hacking and its role in Cybersecurity: A Comprehensive Review

Fatima Asif
231980015
231980015@gift.edu.pk
BS Data Science

Fatima Sohail
231980004
231980004@gift.edu.pk
BS Data Science

Zuhaib Hussain Butt
GIFT University
zuhaib.hussain@gift.edu,pk

Faiz Nasir
231980034
231980034@gift.edu.pk
BS Data Science

Nida Asgar
231980069
231980069@gift.edu.pk
BS Data Science

*Abstract*—This review paper investigates the diverse functions of ethical hacking within modern cybersecurity. By integrating current research, it analyzes the progression of ethical hacking techniques,their use in identifying vulnerabilities and conducting penetration tests, and their influence on strengthening organizational security. Additionally, the paper discusses the ethical considerations, legal contexts and challenges that arises with ethical hacking. This review ultimately enhances the understanding of how ethical hacking can bolster cybersecurity defenses.

## I. INTRODUCTION

This digital environment is marked by an intensifying battle between cybercriminals and those defending against them. Ethicalhacking, which involves proactivaley detecting and addressing vulnerabilities, has become a vital part of cybersecurity efforts.By mimicking real-world cyberattacks, ethical hackers offer essential insights into system vulnerabilities, allowing organizations to enhance their defesnses before they are targeted by malicious actors.This review paper seeks to thoroughly explore the significance of ethical hacking in protecting digital assets. By integrating current research, the study aims to enhance understanding of ethical hacking's effects on organizational security and to highlight potential areas for further investigation.

## II. THE EVOLUTION OF ETHICAL HACKING

Ethical hacking has undergone significant transformation since its early days. Although the term "ethical hacking" was first introduced by IBM Vice President John Patrick in 1995, the concept of using hacking skills for defensive purposes dates back much further.

### A. Early Days: Hacking as a Form of Exploration

Originally, "hacking" was associated with problem-solving and technological innovation. Early hackers, particularly those at institutions like MIT, were motivated by a deep curiosity and a desire to fully understand systems. Their exploratory approach often led to important advancements in both software and hardware

### B. The Emergence of Cybercrime and Ethical Hacking

During the 1980s and 1990s, cybercrime began to rise as hackers increasingly exploited system vulnerabilities for malicious purposes. This change in the hacking landscape created a need for a defensive counterpart. Ethical hacking emerged as this solution, utilizing the same technical skills as malicious hackers but with the aim of safeguarding systems rather than attacking them.

### C. The Digital Revolution and Ethical Hacking

The rise of the internet and growing dependence on digital infrastructure further propelled the evolution of ethical hacking. As networks expanded and became more intricate, the potential for attacks also grew. Ethical hackers adapted by refining their methods to tackle new challenges, concentrating on areas like web application security, network security, and system management.

### D. The Modern Landscape: Ethical Hacking as a Strategic NecessityDigital

In today's world, ethical hacking is recognized as a vital part of cybersecurity strategies. What began as a niche activity has now become a mainstream profession. Organizations of all sizes employ ethical hackers to uncover vulnerabilities, evaluate risks, and strengthen their security measures. Integrating ethical hacking into broader security strategies is now essential for defending against increasingly sophisticated cyber threats.

## III. ETHICAL HACKING APPROACHES AND PRACTICES

Ethical hacking involves utilizing a variety of techniques to identify and exploit security weaknesses. This section offers an in-depth examination of common ethical hacking approaches, such as vulnerability scanning, penetration testing, social engineering, web application hacking, and network hacking. The paper explores the advantages and drawbacks of each approach and their relevance in various organizational settings.

### A. Vulnerability Scanning

Vulnerability scanning is an automated process used to detect potential security flaws in systems or networks. This method employs specialized tools to identify issues like open ports, missing security patches, and outdated software. Although effective in uncovering known vulnerabilities, it often produces a significant number of false positives.

*B. Penetration Testing*

- Penetration testing involves simulating real-world attacks to assess a system's security. By effectively misusing vulnerabilities, it assesses an organization's capacity to identify and react to dangers. Penetration testing can be performed in black box, white box, or gray box scenarios, depending on the level of information available to the tester

*C. Social Engineering*

Social Engineering controls human brain research to betray people into uncovering delicate data or performing activities that compromise security.. This method typically involves techniques like phishing, pretexting, and baiting. Despite being difficult to defend against, its impact can be minimized through awareness training and strong security policies.

*D. Web Application Hacking*

Web application hacking targets vulnerabilities within web applications, using techniques such as SQL injection, cross-site scripting (XSS), cross-site request forgery (CSRF), and session hijacking. Safeguarding web applications requires secure coding practices, input validation, and routine security assessments.

*E. Network Hacking*

Network hacking focuses on exploiting vulnerabilities in network infrastructure, including routers, switches, and firewalls. Methods like sniffing, spoofing, and denial-of-service (DoS) attacks are commonly used. Effective countermeasures include network segmentation, intrusion detection systems, and the use of firewalls.

IV. ETHICAL HACKING IN VULNERABILITY ASSESSMENT AND MANAGEMENT

Vulnerability assessment plays a vital role in any cybersecurity strategy. This section delves into how ethical hacking contributes to identifying and prioritizing vulnerabilities, as well as in crafting effective remediation strategies. The discussion covers the integration of ethical hacking within vulnerability management frameworks and emphasizes the necessity of ongoing vulnerability assessments to maintain robust organizational security.

*A. The Role of Ethical Hacking in Vulnerability Assessment*

Ethical hacking offers a proactive and dynamic method for vulnerability assessment. By replicating real-world attack scenarios, ethical hackers can identify vulnerabilities that automated tools might overlook.This cooperative energy between human skill and innovative devices improves the by and large viability of the defenselessness appraisal handle.

*B. Prioritizing Vulnerabilities*

Ethical hackers play a crucial role in prioritizing vulnerabilities based on their potential impact on the organization. By evaluating the likelihood of exploitation and the potential damage of a successful attack, they help organizations adopt a risk-based approach to vulnerability management.

*C. Crafting Remediation Strategies*

Ethical hackers provide valuable insights for developing robust remediation strategies. By thinking like attackers, they can suggest countermeasures that are more effective in preventing exploitation. Furthermore, they can perform retesting to ensure that implemented patches and controls are effective.

*D. Integrating Ethical Hacking with Vulnerability Management Frameworks*

To fully leverage ethical hacking, it should be integrated into a comprehensive vulnerability management framework. This integration involves aligning ethical hacking activities with other security processes, such as asset management, risk assessment, and incident response. Establishing a continuous feedback loop between ethical hacking and vulnerability management strengthens an organization's security posture.

*E. The Necessity of Continuous Vulnerability Assessment*

Given the constantly evolving threat landscape, with new vulnerabilities frequently emerging, continuous vulnerability assessment is crucial. Supported by ethical hacking, ongoing assessments help organizations stay ahead of potential attackers

V. ETHICAL HACKING AND PENETRATION TESTING

Penetration testing involves simulating an attack on a computer system or network to assess its security.

This section explores the role of ethical hacking in conducting penetration tests, covering different types of testing (black box, white box, gray box) and the methodologies used. The discussion emphasizes the importance of designing, executing, and reporting tests effectively to achieve successful penetration testing outcomes.

A. Types and Methodologies of Penetration Testing.

Penetration testing can be conducted using various approaches, each with distinct strengths and weaknesses:

- Black-box testing: The tester has no prior knowledge of the system, mimicking a real-world attack. This method can reveal vulnerabilities that might be missed in other testing approaches.

- White-box testing: The tester has full knowledge of the system, including network diagrams, source code, and user credentials, enabling a thorough examination of vulnerabilities.

- Gray-box testing: The analyzer has halfway information of the framework, comparative to an insider with constrained get to. This approach combines the points of interest of both black-box and white-box testing.

Penetration testing methodologies generally follow these phases:

1. Planning and reconnaissance: Establishing test objectives, identifying the target system, and gathering relevant information about the environment.

2. Scanning and discovery: Using automated tools and manual techniques to detect potential vulnerabilities.

3. Abuse: Endeavoring to misuse distinguished vulnerabilities to pick up unauthorized get to.

4. Post-exploitation: Assessing the impact of the attack by exploring the compromised system.

5. Reporting: Documenting the findings, offering recommendations, and outlining remediation steps.

B. Importance of Test Design, Execution, and Reporting

Effective penetration testing requires thorough planning and execution. A well-crafted test arrange characterizes the scope, targets, and technique of the evaluation. The testing phase should be carried out systematically, adhering to established procedures and ethical standards. Detailed reporting is essential for conveying test results to stakeholders and aiding in the remediation process.

VI. Ethical Hacking and Penetration Testing

Ethical Principles in Ethical Hacking:

Ethical hacking is founded on a set of core principles that guide practitioners in their work:

- Authorization: Moral programmers must secure express consent from framework proprietors some time recently performing any exercises. This authorization ought to be formally reported in an understanding that clearly traces the scope and confinements of the engagement.

- Non-disruption: Ethical hackers are responsible for ensuring that their activities do not cause harm or disrupt systems or networks. This includes safeguarding against unauthorized access to sensitive data and avoiding any interruptions in service.

- Confidentiality: Any information gathered during ethical hacking must be kept confidential. Ethical hackers are obligated to protect sensitive data and intellectual property from unauthorized disclosure.

- Legality: Ethical hacking practices must always adhere to relevant laws and regulations. This includes understanding and following data protection laws, privacy regulations, and cybersecurity frameworks.

Legal Frameworks Governing Ethical Hacking

- The legal environment surrounding ethical hacking is intricate and varies by jurisdiction. Important legal considerations include:

- Licensing and certification: In some regions, ethical hackers are required to obtain specific licenses or certifications. These credentials can help validate their expertise and offer legal protection.

- Liability: Ethical hackers could face legal liability if their actions result in harm or damage. Therefore, clear contractual agreements are crucial to define roles, responsibilities, and limitations.

- Data protection: Ethical hackers must handle personal and sensitive data with care, complying with data protection laws such as GDPR or CCPA.

- Computer crime laws: These laws define the legal limits of hacking activities and can differ widely between countries. Ethical hackers need to be well-versed in the specific laws relevant to their operations.

VII. Ethical Principles in Ethical Hacking

The Impact of Ethical Hacking on Organizational Security:

Ethical hacking is crucial for strengthening organizational security. By identifying vulnerabilities before they can be exploited by malicious actors, ethical hacking supports a proactive security stance. When combined with other security measures, it forms a comprehensive defense-in-depth strategy.

Enhancing Organizational Resilience

- Incident Response: Ethical hacking provides critical insights into potential attack vectors, allowing organizations to craft effective incident response plans. By simulating real-world attacks, ethical hackers can uncover flaws in current incident response procedures and suggest necessary improvements.

- Security Awareness Training: Results from ethical hacking assessments can inform the development of focused security awareness training programs. Understanding the types of attacks that might succeed enables organizations to educate employees on common social engineering tactics and best practices for safeguarding sensitive data.

- Risk Management: Ethical hacking aids organizations in accurately identifying and prioritizing risks. By evaluating the potential impact of vulnerabilities, organizations can effectively allocate resources to mitigate the most critical risks.

VIII. Challenges and Future Directions

Despite its importance, ethical hacking faces several significant challenges:

- Skill Shortage: There is a global lack of skilled ethical hackers, making it challenging for organizations to build strong, effective teams.

- Evolving Threat Landscape: The fast pace of technological advancements and the rise of new threats

demand continuous updates and adaptations of ethical hacking techniques.

- Ethical Dilemmas: Balancing the need to uncover vulnerabilities while protecting sensitive information can lead to ethical challenges.

To overcome these challenges, future research should concentrate on:

- Developing Automated Ethical Hacking Tools: Utilizing AI and machine learning to improve the efficiency and effectiveness of ethical hacking practices.

- Ethical Hacking Education and Training: Expanding training programs to cultivate a more skilled workforce in ethical hacking.

- Ethical Frameworks: Establishing clear guidelines to ensure ethical conduct in ethical hacking activities.

Moreover, integrating ethical hacking with emerging technologies such as cloud computing, IoT, and blockchain is crucial for tackling future cybersecurity challenges.

## IX. GENERAL ANALYSIS:

### A. Distribution of Cybersecurity Incidents Prevented by Ethical Hacking

Figure 1 shows the distribution of various types of cybersecurity incidents that were prevented due to ethical hacking. The chart indicates that a significant portion of prevented incidents were data breaches, followed by ransomware attacks.

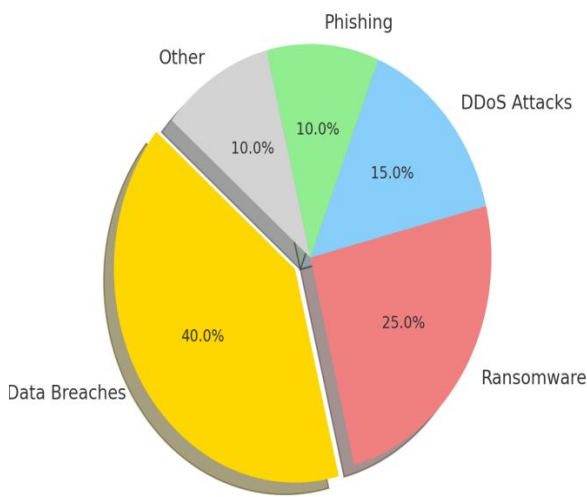

Figure 1: Distribution of various types of cybersecurity Incidents

### B. Growth in Ethical Hacking Adoption Over the Years

This bar graph illustrates the increase in the adoption of ethical hacking practices from 2015 to 2024. The data shows a steady rise, reflecting the growing awareness and necessity of ethical hacking in cybersecurity.

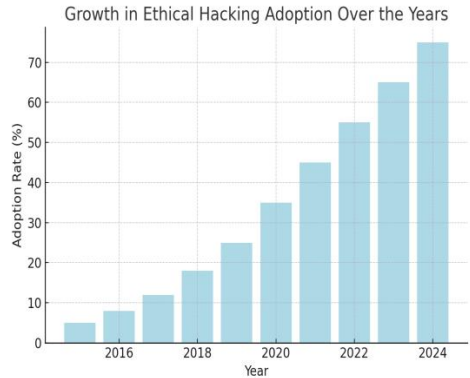

Figure 2: Increase in the adoption of ethical hacking practices from 2015 to 2024

### C. Comparison of Security Measures

The following table compares the effectiveness of different security measures, including ethical hacking, antivirus software, firewalls, and security audits. Ethical hacking is distinguished by its proactive strategy in uncovering security weaknesses.

| Security Measure | Proactive Defense | Cost | Implementation Complexity | Overall Effectiveness |
|---|---|---|---|---|
| Ethical Hacking | High | Moderate | High | Very High |
| Antivirus Software | Low | Low | Low | Moderate |
| Firewalls | Medium | Low | Moderate | High |
| Security Audits | Medium | Moderate | High | High |

### D. The Impact of Ethical Hacking on Reducing Security Vulnerabilities

Figure 3 compares the number of security vulnerabilities before and after the implementation of ethical hacking practices. The data suggests a significant reduction in vulnerabilities, highlighting the effectiveness of ethical hacking.

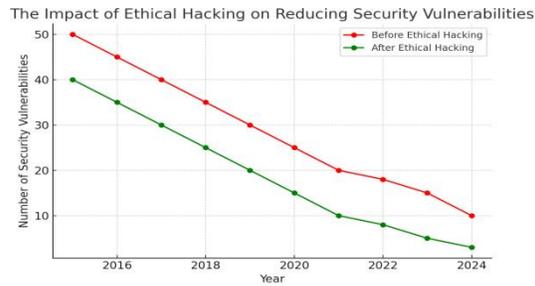

Figure 3: Comparison of number of security vulnerabilities

CONCLUSION

Ethical hacking has ended up an fundamental component of advanced cybersecurity. By proactively identifying and addressing vulnerabilities, ethical hackers are crucial in protecting organizations from the continuously evolving threat landscape. This review has highlighted the diverse aspects of ethical hacking, including its evolution, methodologies, application in vulnerability assessment and penetration testing, and its broader impact on organizational security.

Despite its many advantages, ethical hacking also presents challenges and ethical dilemmas. Overcoming these issues will require collaborative efforts from policymakers, industry experts, and researchers. As technology progresses, the role of ethical hacking is expected to grow, making ongoing research and development vital to staying ahead of new threats.

In summary, ethical hacking is more than just a defensive measure; it is a strategic necessity for organizations aiming to safeguard their digital assets. By cultivating a security-conscious culture and investing in skilled ethical hackers, organizations can greatly enhance their resilience against cyberattacks and secure a safer digital future.